# Exchange Rate Sensitivity in Free Zone Trade: An Empirical Study of the Istanbul Atatürk Airport Free Zone

Şükrü C. DEMİRTAŞ[1]

March 5, 2025

**Abstract**

This study as part of an ongoing research effort, empirically examines the relationship between foreign trade in the Istanbul Atatürk Airport Free Zone and exchange rate movements. Monthly data from 2003 to 2016 were analyzed through stationarity tests, followed by the Vector Autoregressive (VAR) model, Cointegration Analysis, and the Toda-Yamamoto Causality Test. The findings indicate that the exchange rate does not significantly affect imports and exports in the free zone. This result suggests that free zones, due to their structural characteristics and operational framework, may be relatively insulated from exchange rate fluctuations. The study contributes to the literature by providing a focused analysis of a specific free zone in Türkiye, highlighting the potential independence of free zone trade from exchange rate volatility.

[1] Graduate School of Social Sciences, Yildiz Technical University, Istanbul, Turkey, OID: 0000-0002-6953-3599, demirtasukru@gmail.com

## 1. Introduction

As the globalization effort of the world economies and the specialization in production accelerated by this effort increase, the mutual demand of countries for each other's goods also increases. The realization of the positive impact of globalized international trade and exchange rates on countries has led researchers to search for new ways to be more profitable in foreign trade. In this context, countries that want to benefit more from the benefits of international trade and to improve their welfare level, to sustain the development of the country and to realize the domestic production of industrial products imported from abroad increasingly need foreign currency. For this reason, foreign trade needs to be increased. However, each country imposes restrictions to protect the domestic market and producers through various agreements. Considering that the possibilities of finding foreign credits are very limited, countries have to concentrate on all kinds of foreign exchange earning transactions and earn foreign exchange in the short term. For this reason, a new area that is not affected by the trade restrictions mutually imposed by countries is required and in this context, free zones come to the agenda (Bağrıaçık, 1983).

There are various definitions of free zones. In the report submitted to the Ministry of Trade by the Istanbul Chamber of Commerce, it is stated that "Free Zones are not only places consisting of one or many buildings where goods of the final product nature will be kept, but also places where it is possible to keep a whole area outside the customs borders, and counting wider activities in a certain area within the national borders outside the customs border in order to carry out activities without being subject to tax payment obligation constitutes the establishment called free zone" (ITO, 1960).

Since free zones are established to provide countries with positive effects in foreign trade and commercial advantages, these effects are expected to provide gains in international markets. In general, free zones have positive effects on employment, foreign exchange accumulation, imports and foreign capital inflow; however, there are also negative effects such as tax loss, import pressure, risk of increase in illegal activities, smuggling, rivalry with domestic firms, negative distribution of investments, monopolization and deterioration of social structure. Free zones also offer various benefits to producers, including tax advantages, the ability to plan for the medium and long term, profit transfer, trade facilitation, exemption from customs duties, free movement aligned with European Union and customs union criteria, the principle of equality, no time restrictions, market-driven commercial flexibility, realistic inflation accounting, access to both domestic and international markets, streamlined bureaucratic procedures, dynamic business management, strategic advantages, cost-effective infrastructure suitable for diverse commercial and industrial activities, and improved access to supply chain facilities (Ekonomi bakanlığı, 2016).

Free zones are zones established to export production using external intermediate inputs as well as factors of production in a country. The resulting payments and expenditures are made at the exchange rate, calculated at the exchange buying rate of the central bank on that day. From this point of view, exports and imports in free zones are affected by changes in the exchange rate. Because production is carried out not only by using imported inputs, but also by using many domestic intermediate inputs such as labor and capital, natural resources and so on. The costs of these inputs also change with the changes in exchange rates.

Although the exchange rate effect is considered to be relatively small, this study was conducted to examine its effects. Since free zones trade in exchange rates, the effects of exchange rate changes can be analyzed through the terms of trade of free zones. In this study, it will be investigated whether the advantages provided by free zones in foreign trade, which are relatively less affected by global events, are affected by exchange rate changes, which are also related to foreign trade. For the originality of the study, the relationship between the foreign trade components of Istanbul Atatürk Airport Free Zone in Türkiye and the real effective exchange rate is analyzed. The study is organized as follows: Section 2 related literature, Section 3 methodology, Section 4 results, and Section 5 concludes.

## 2. Related literature

In the research conducted, it is seen that there are few studies on free zones and exchange rate. In addition to these studies, there are also studies that examine the relationship between free zones and other variables (such as foreign trade relations, imports and exports and their price indices).

Terzi and Zengin (1999), using VAR analysis, examined the role of the relationship between exchange rate and foreign trade variables in achieving foreign trade balance for Türkiye for the years 1989-1996 and found that the necessary conditions for Türkiye's exchange rate movements to positively affect foreign trade have not yet been met and the results support the hypothesis that the exchange rate effect cannot be an effective tool in achieving foreign trade balance.

Zengin (2001) examined the validity of the reflection and standard theory approach for the Turkish economy for the period 1994-2000. The study employed cointegration and VAR analysis methods, revealing a long-term cointegration relationship among the import price index, export price index, and real exchange rates. Additionally, the VAR analysis results indicate a causal relationship from the indices to real exchange rates.

Gürbüz and Çekerol (2002) examined the relationship between foreign trade volume, import price index, export price index variables and exchange rate separately using monthly data for the period 1995-2002 and analyzed that there is no relationship between the variables.

Karagöz and Doğan (2005) analyzed monthly data from 1995 to 2004, first examining the stationarity of the series for cointegration. Their findings indicate that the series were stationary and that there was no direct one-to-one economic relationship between the exchange rate and either imports or exports. However, they found that the impact of the 2001 devaluation was significant.

Gül and Ekinci (2006) examined the relationship between real exchange rates, exports, and imports in Türkiye from 1990 to 2006 using the Granger causality test. Their findings indicate no causal relationship from real exchange rates to exports or imports. However, they identified a unidirectional causality from exports and imports to real exchange rates.

Ay and Özşahin (2007) tested the J curve hypothesis for Türkiye (1995-2007) using time series analysis. The Augmented Dickey-Fuller test confirmed stationarity, and Granger causality analysis showed that while the real exchange rate and export price index influence the import price index, only the real exchange rate affects the export price index. The real exchange rate was found to be independent of both variables.

Gerni et al. (2008), using the Feder 1982 model, find that exports are an important determinant of economic growth in growth equation estimations for the period 1980-2006.

Tyler and Negrete (2009) employed panel data analysis to compare the economic growth of 87 countries from 1961 to 1999, distinguishing between countries that adopted the free zone system and those that did not. The findings indicate a positive impact of free zone implementation on economic growth, with countries utilizing free zones exhibiting faster economic growth compared to those without such systems.

Hao (2010) investigates the effects of free zones on growth in Shanghai economy for the period 1996-2009. According to the findings obtained by Granger Causality analysis with annual data, it is concluded that free zones contribute to growth.

Yapraklı (2010) examined the effects of budget deficit, real money supply, and real effective exchange rate index on the real foreign trade deficit using monthly data from 2001 to 2009. The results show that, in the long run, the foreign trade deficit is positively and statistically significantly influenced by the budget deficit, while money supply has a negative but statistically insignificant effect. In the short run, money supply has a positive and significant impact on the trade deficit. Both the budget deficit and real effective exchange rate index exhibit parallel effects in the short and long run.

Karaçor and Gerçeker (2012) analyzed the relationship between real exchange rates and foreign trade in Türkiye from 2003 to 2010. Their findings revealed a cointegration relationship between real exchange rates and foreign trade volume, as well as a causality from real exchange rates to foreign trade volume in both the short and long run. Additionally, they found a causality from foreign trade volume to real exchange rates only in the short run.

Yıldırım and Kesikoğlu (2012) examined the causality between Türkiye's imports, exports, and real exchange rate from 2003 to 2011 using the MWALD test with the leveraged bootstrap technique by Hacker and Hatemi-J (2006). They found bidirectional causality between total exports and imports, as well as between exports/imports of intermediate and capital goods. A unidirectional causality was observed from total exports to consumption goods imports and from consumption goods exports to both imports of consumption goods and intermediate goods. No causality was found between exchange rates and imports/exports. The results suggest that the interdependence of imports and exports makes exchange rate policy unbiased.

Yılmaz and Çapraz (2014) tried to explain the efficiency of the regions by using data on 12 free zones in Türkiye with the help of data envelopment analysis and concluded that 8 of the regions were efficient.

Yıldırım and Kılıç (2014) investigated the effects of exchange rate volatility on Türkiye's exports to Euro Area countries using panel data analysis for the period 2000-2012. They also examined this effect at different time frequencies by applying discrete wavelet transform on the data. Their findings suggest that exchange rate volatility does not have a negative impact on Türkiye's exports to Euro Area countries in the short run, but may have negative effects on exports in the long run.

Öncel and Demirtaş (2016) analyzed whether the exports and imports of free zones in Türkiye affected Türkiye's foreign trade in the 2000-2016 period by using Hacker and Hatemi-J causality tests. They found a causality relationship from Türkiye's exports to free zones' exports, from free zones' exports to Türkiye's exports, from Türkiye's imports to free zones' imports, from free zones' imports to Türkiye's imports, from Türkiye's total foreign trade to free zones' total foreign trade. At the same time, they did not find a causality relationship from the total trade volume of free zones to the total volume of Türkiye.

Öncel and Demirtaş (2017) analyzed the effect of trade from free trade zones on Türkiye's foreign trade using ARDL test and monthly data for 2000-2015. It is found that there is both a short-run and a long-run relationship between the variables.

Bicil and Uçak (2018) investigated the relationship between exports from free zones and total exports for the period 1988-2017 for Türkiye. ARDL bounds test analysis was conducted and a cointegration relationship was found between the series. In the long run, a 1% increase in exports from free zones will increase total exports by 0.86%, while in the short run, a 1% increase in exports from free zones will increase total exports by 0.22%. It is concluded that exports from free zones make a positive contribution to Türkiye's total exports in the long run

Aktepe and Taştan (2019), in their study aiming to measure the foreign trade efficiency of free zones in Türkiye between 2003-2018 according to certain product groups and to group them among themselves, concluded that the classification of free zones in Türkiye according to certain product groups with clustering analysis and the import or export of products with high foreign trade performance from free zones will increase the foreign trade performance of free zones.

Li et al. (2021) investigated the impact of free zones on the performance of port enterprises on 16 port companies in China by conducting panel data analysis for the period covering the years 2010-2016. They concluded that the effect of free zones on the scale of companies, labor productivity and management efficiency is positive.

Avcı et al. (2022) tried to explain the efficiency of free zones for the Covid pandemic period. In their study, conducted a comparative analysis by estimating two separate models for the years 2019-2020 and concluded that although Türkiye's foreign trade volume decreased during the pandemic period, the foreign trade volume of free zones increased. In addition, the effect of free zones on employment increased during the pandemic period.

Ateş et al. (2023), in their study investigating the contribution of firms in free zones to Türkiye's foreign trade and employment and the relationship between these variables, conducted Fourier ADL hidden cointegration test with 7 different models with variables determined with monthly data for 2013-2022. It is concluded that the contribution of imports and exports of free zones with tax advantages to Türkiye's foreign trade and employment mechanism is weak.

Duran (2023), free zones are ranking analyzed with WASPAS and GIA methods by calculating their weights with CRITIC method. 18 free zones were analyzed with 2022 data. According to the findings, it is determined that the factor that predominantly affects free zones is the import variable. In terms of export performance, Aegean Free Zone ranks first, Rize Free Zone ranks last. As a result, it is concluded that free zones make significant contributions to Türkiye's foreign trade.

Bayraktutan and Acar (2023) utilized panel data analysis to investigate the factors influencing the export performance of 18 free zones in Türkiye from 2004 to 2020. The model was estimated using the Driscoll-Kraay estimator, and the Dumitrescu-Hurlin causality test was applied. Their findings indicate that the logistics capacity of the province where the free zone is situated, the level of education, the exchange rate, and the average economic growth of EU countries have a positive impact on exports from free zones.

İmamoğlu (2024) examined the activities of free zones established in Türkiye by taking into account the establishment objectives of free zones, the level of realization of these objectives and the effectiveness of the zones in directing the economy.

Sun et al. (2025), Drawing on a quasi-natural experiment involving pilot free trade zone policies, this study utilizes panel data from China's A-share listed companies between 2008 and 2022 to empirically examine the impact of these policies on firms' green transformation. Using a multi-period difference-in-differences approach, the findings reveal that pilot free trade zone policies significantly enhance firms' green transformation efforts. An effect test notes that these policies affect enterprises' green transformation via financing constraints and green innovation level. The heterogeneity test results indicate that, compared to enterprises and state-owned enterprises (SOEs) in the central and western regions, the impact of these policies on green transformation is more pronounced for enterprises and non-SOEs in the eastern region.

## 3. Methodology

This study analyzes whether exports and imports of Istanbul Atatürk Airport Free Zone are affected by the real effective exchange rate with 157 monthly data between 2003 and 2016. First, the data are seasonally adjusted. Then, stationarity test was performed, VAR model was constructed, followed by cointegration analysis and Toda Yamamoto (1995) causality test. The data used in the study were obtained from the Turkish Statistical Institute and Istanbul Atatürk Airport Free Zone Customs Directorate.

### 3.1. Stationarity Test

Stationarity means that the series approach a certain value or fluctuate around the expected value (Bozkurt, 2007). In order to obtain accurate results from time series, the series should be stationary. In analyzes conducted with non-stationary time series, the problem of spurious regression arises (Granger and Newbold ,1974). For this reason, the results obtained cannot reflect the reality. For this reason, the series must first be stationary.

If the mean, variance and covariance of a time series do not change over time, it is defined as weak stationarity (Darnel, 1994). If the series are stationary in their first differences, the series is called first degree integrated (Kennedy et al., 2006). In this study, the Augmented Dickey-Fuller (ADF) Unit Root Test is used. The test results are shown in Table 1.

### 3.2. Cointegration Test

Cointegration implies the existence of a long-run relationship between series. According to the cointegration theory, it is only possible to return to equilibrium when there is a deviation from equilibrium or for deviations from equilibrium to be temporary with cointegration relationship. Therefore, if non-stationary variables are cointegrated, it is not appropriate to take the differences of the variables. Because these variables have a common trend that moves together. Taking the difference eliminates the common trend and leads to loss of statistical information. In this study, Johansen Cointegration test was conducted. The test results are shown in Table 5.

### 3.3. Causality Test

In the Toda Yamamoto (1995) causality test, VAR is used with level variables and these models can investigate the dynamic relationships between variables. In addition, the presence of a possible cointegration relationship between variables in this method does not change the results of the analysis (Tapşin and Karabulut, 2013). This test does not require the tests used in the determination of unit root and cointegration properties, and the risk of incorrectly determining the degree of integration of the series is minimized. The test results are shown in Table 6.

## 4. Results

In order to evaluate the relationship between the variables properly, the series should be stationary, that is, they should not contain unit root. In this context, the Augmented Dickey-Fuller (ADF) Unit Root Test was conducted for stationarity test.

**Table 1: ADF Unit Root Test Results**

| Variables | T Statistics | T Statistic (I.Difference) |
|---|---|---|
| DK | 2.1041 (0.9999) | -8.8257 (0.0000) |
| IHR_SA | -1.1715 (0.6860) | -14.4475 (0.0000) |
| ITH_SA | -1.3907 (0.5856) | -13.0917 (0.0000) |

Note: Values in parentheses denote probability (p) values.

Table 1 displays the ADF Unit Root Test for the variables DK, IHR_SA, and ITH_SA in both their original form and first differences. The results show that the variables are not statistically significant at level but become significant when first differences are applied. This suggests that their changes over time have a meaningful impact, which allows for further analysis, such as VAR and cointegration, using differenced data. For the VAR model, the appropriate lag length should be determined first.

**Table 2: VAR Lag Length**

| LL | LogL | LR | FPE | AIC | SC | HQ |
|---|---|---|---|---|---|---|
| 0 | -3363.74 | NA | 8.49e+15 | 45.19 | 45.25 | 45.21 |
| 1 | -2903.26 | 896.23 | 1.98e+13 | 39.13 | 39.37 | 39.22 |
| 2 | -2878.49 | 47.22 | 1.60e+13 | 38.91 | 39.34* | 39.09 |
| 3 | -2859.55 | 35.32* | 1.40e+13* | 38.78* | 39.39 | 39.03* |
| 4 | -2855.13 | 8.08 | 1.49e+13 | 38.84 | 39.63 | 39.16 |
| 5 | -2847.11 | 14.30 | 1.52e+13 | 38.86 | 39.82 | 39.25 |
| 6 | -2840.13 | 12.18 | 1.56e+13 | 38.88 | 40.03 | 39.35 |
| 7 | -2837.45 | 4.55 | 1.70e+13 | 38.97 | 40.30 | 39.51 |
| 8 | -2835.08 | 3.94 | 1.87e+13 | 39.06 | 40.57 | 39.67 |

Note: LL: Lag Length, LR: Jointly Adjustable LR Test Statistic, FPE: Final Prediction Error, AIC: Akaike Information Criterion, SC: Schwarz Information Criterion, HQ: Hannan-Quinn Information Criterion. The asterisks (*) denote the lag lengths that minimize the AIC, SC, and HQ criteria, confirming the selection of lag 3 as the optimal model.

Table 2 presents the results of model selection criteria, including Log-Likelihood (LogL), likelihood ratio (LR), Final Prediction Error (FPE), Akaike Information Criterion (AIC), Schwarz Criterion (SC), and Hannan-Quinn Criterion (HQ) for different lag lengths (0 to 8). The table indicates that as the lag length increases, the values of LogL improve, but the AIC, SC, and HQ criteria show a decline up to a lag length of 3, with lag 3 yielding the lowest values for AIC (38.78), SC (39.39), and HQ (39.03). These results suggest that the optimal lag length is 3, as it minimizes the AIC, SC, and HQ, providing the most appropriate model for further analysis. To see whether this model is stable, the inverse roots of the AR characteristic polynomial should be examined

**Table 3: Inverse Roots of AR Characteristic Polynomial**

| Root | Modulus |
|---|---|
| 0.988076 | 0.988076 |
| 0.883526 | 0.883526 |
| 0.833718 | 0.833718 |
| 0.184852 - 0.518429i | 0.550398 |
| 0.184852 + 0.518429i | 0.550398 |
| -0.256203 - 0.429323i | 0.499959 |
| -0.256203 + 0.429323i | 0.499959 |
| -0.207442 - 0.309839i | 0.372870 |
| -0.207442 + 0.309839i | 0.372870 |

Source: Created by the author.

Table 3 presents the inverse roots of the AR characteristic polynomial. The modulus values of all roots are below 1, which indicates that the VAR model is stable. Additionally, the stability of the model is confirmed by the position of the inverse roots within the unit circle, as all roots lie inside it. This suggests that the chosen lag length results in a stable model, suitable for further analysis.

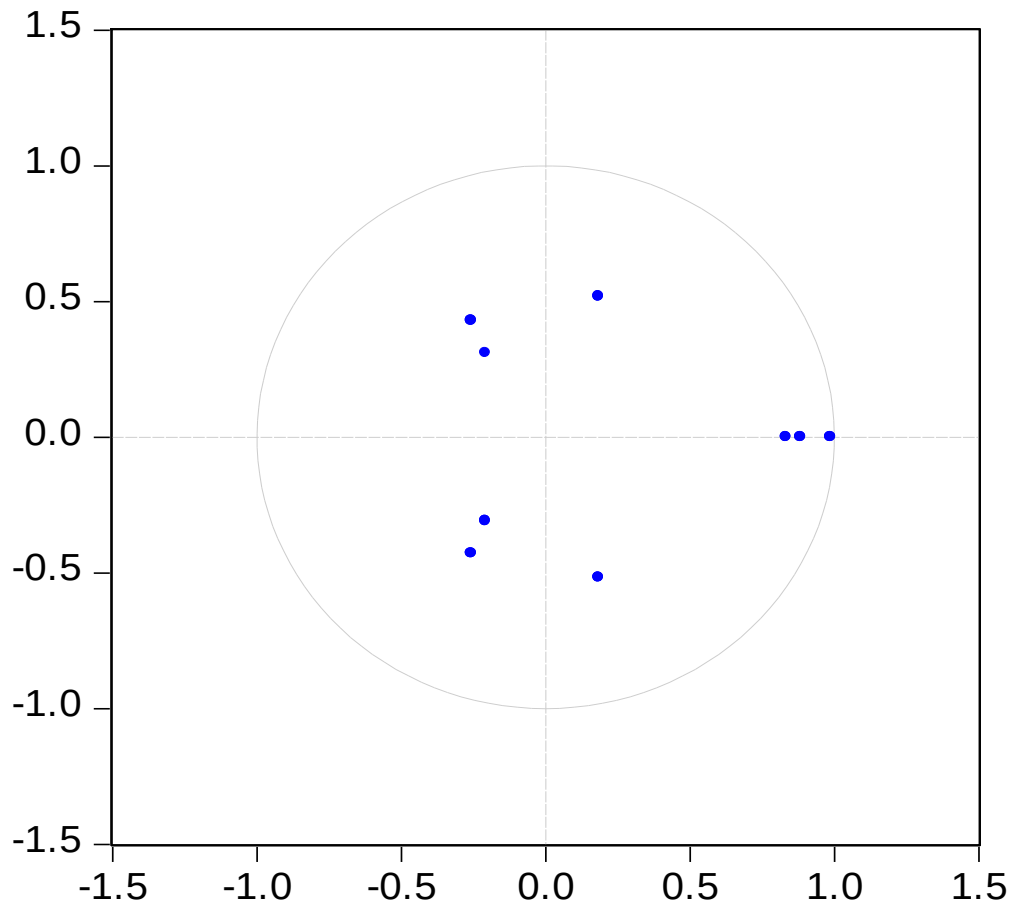

**Figure 1: Unit Circle Location of the Inverse Roots of the AR Characteristic Polynomial**

Figure 1 illustrates the unit circle positions of the inverse roots of the AR characteristic polynomial. As all AR roots lie within the unit circle, it confirms that the model is stationary. To further assess the model, an Autocorrelation LM test is conducted to check for interdependence between the series. If interdependence exists, it would indicate that the model is statistically insignificant and structurally inconsistent.

**Table 4: Autocorrelation LM Test**

| Lag | LM Statistics | Probability |
|---|---|---|
| 1 | 9.5480 | 0.3883 |
| 2 | 9.2779 | 0.4120 |
| 3 | 13.084 | 0.1588 |
| 4 | 9.0456 | 0.4331 |
| 5 | 8.4861 | 0.4860 |
| 6 | 8.9920 | 0.4380 |
| 7 | 4.1472 | 0.9015 |
| 8 | 7.1348 | 0.6231 |
| 9 | 6.7195 | 0.6663 |
| 10 | 11.344 | 0.2528 |
| 11 | 5.6996 | 0.7696 |
| 12 | 4.9943 | 0.8348 |

Source: Created by the author.

Table 4 presents the results of the Autocorrelation LM test, showing the LM statistics and corresponding p-values for each lag. Since all p-values are greater than the conventional significance level (0.05), there is no evidence of interdependence between the series at any lag. This indicates that the model is statistically significant and structurally consistent, as no significant autocorrelation is detected among the residuals. After the findings, it is concluded that the cointegration test can be performed.

The Johansen cointegration test was conducted to determine whether the import-export and exchange rate series used in the study are cointegrated. The test helps to identify the presence of a long-term relationship between the series, indicating that they move together over time despite potential short-term fluctuations. The results of the cointegration test will provide further insights into the dynamics between the import-export and exchange rate series.

**Table 5: Johansen Cointegration Test**

| Hypothesis | Eigenvalue | Trace Statistics | 0.05 Critical Value | Probability |
|---|---|---|---|---|
| r=0 | 0.0701 | 20.1646 | 35.1927 | 0.7164 |
| At most r=1 | 0.0438 | 9.0433 | 20.2618 | 0.7322 |
| Mostly r=2 | 0.0141 | 2.1796 | 9.1645 | 0.7417 |

Note: r denotes the number of cointegration vectors.

Table 5 presents the results of the Johansen cointegration test, which examines whether there is a long-term relationship between the import-export and exchange rate series. Based on the Schwarz information criterion, the second model, which includes a constant but no trend and does not include a constant in the VAR model, is considered the appropriate model. However, the results indicate no cointegration between the series. According to the Trace Statistics, all values are well above the 0.05 significance level, and the probability values are high, suggesting no cointegration between the series. Therefore, it is concluded that no cointegration exists between these variables. In light of this, the Toda Yamamoto causality test, which does not require cointegration, will be performed next.

The data used in the study were subjected to Toda Yamamoto (1995) causality test to determine the direction of the causality relationship. The results of the tests where exchange rate is the dependent variable are shown in Table 6.

**Table 6: Toda Yamamoto (1995) Causality Test**

| **Dependent Variable** | **Explanatory Variables** | | |
|---|---|---|---|
| | IHR_SA | DK | ITH_SA |
| IHR_SA | | 3.2017 (0.5246) | |
| ITH_SA | | 3.9824 (0.4084) | |

Note: Values in parentheses indicate probability values, while values outside parentheses indicate chi-square statistics.

Table 6 presents the results of the Toda Yamamoto (1995) causality test, with the exchange rate as the dependent variable and IHR_SA, DK, and ITH_SA as the explanatory variables. The chi-square statistics and their corresponding probability values are displayed in the table. According to the test results, there is no significant causality from the exchange rate to the explanatory variables, as the probability values are above the conventional 0.05 significance level. Therefore, based on the Toda Yamamoto causality test, no significant causal relationship is found from the exchange rate to exports and imports.

## 5. Conclusion

Countries such as Türkiye, which are classified as developing countries, require a continuous and increasing amount of capital to finance their development. Therefore, countries are trying to increase their foreign exchange earnings and to establish superiority in foreign trade. Free trade zones are a good alternative to provide this capital.

In conclusion, the results from the various tests conducted in this study, focused on the Atatürk Airport Free Zone, provide valuable insights into the relationships between the exchange rate, imports, and exports. The ADF Unit Root Test confirmed that the series are non-stationary at level but become stationary when first differences are applied, allowing for the use of VAR and cointegration analysis. The VAR model selection criteria indicated that lag 3 is optimal, minimizing the AIC, SC, and HQ criteria. The stability of the model was confirmed through the analysis of the inverse roots of the AR characteristic polynomial, with all roots lying within the unit circle, ensuring the model's stationarity. The Autocorrelation LM test showed no significant interdependence between the series, further confirming the statistical significance and structural consistency of the model. The Johansen cointegration test revealed that there is no cointegration between the exchange rate and import-export series, as all trace statistics were above the 0.05 significance level. Consequently, the Toda Yamamoto causality test was employed, but it also indicated no significant causality from the exchange rate to exports and imports.

Overall, the findings suggest that while no cointegration exists between the exchange rate and trade variables in the Atatürk Airport Free Zone, the lack of causality points to the need for further investigation into the dynamics between these variables using alternative methods or different model specifications.

**Funding**

The author report that they have no funding to declare.

**References**

Aktepe, C., & Taştan, O. C. (2019). *Serbest Bölgelerin Dış Ticarete Etkisinin Veri Madenciliği İle Ölçülmesi ve Sektörel Bir Uygulama*. İşletme Araştırmaları Dergisi, 11(4), 2396-2411.

Ateş, M. S., Ateş, G., Toktaş, D., & Gökçe, E. C. (2023). *Serbest bölgelerdeki ithalat-ihracat ile Türkiye'nin dış ticaret ve istihdam ilişkisi: Fourier ADL saklı eşbütünleşme testi*. İstanbul İktisat Dergisi, 73(1), 385-418.

Avcı, Z., Kovacı, S., & Şen, S. (2022). *Türkiye'de Covıd-19 Döneminde Serbest Bölgelerin Dış Ticaret Ve İstihdam Açısından Önemi*. Fırat Üniversitesi Sosyal Bilimler Dergisi, 32(2), 563-578.

Ay, A., & Özşahin, Ş. (2007). *J Eğrisi Hipotezinin Testi: Türkiye Ekonomisinde Reel Döviz Kuru ve Dış Ticaret Dengesi İlişkisi*, Uludağ Üniversitesi İİBF Dergisi, 26(1).

Bağrıaçık, A. (1983). *Türkiye'de Serbest Bölge Uygulaması.* Yeni İş Dünyası, 47.

Bayraktutan, Y., & Acar, M. (2023). *Serbest Bölgelerin İhracat Performansını Belirleyen Etmenler: Türkiye Örneği*. Uşak Üniversitesi Sosyal Bilimler Dergisi, 16(2), 58-71.

Bicil, İ. M., & Uçak, S. (2018). *Toplam ihracat-serbest bölgeler ihracatı ilişkisi: Türkiye örneği*. Gümrük ve Ticaret Dergisi, (12), 50-63.

Bozkurt, H. (2007). *Zaman Serileri Analizi*, Ekin Kitap Evi, Bursa.

Duran, G. (2023). *Türkiye'de Serbest Bölgelerin Performanslarının CRİTİC Tabanlı WASPAS ve GİA Uygulamaları ile Değerlendirilmesi.* MANAS Sosyal Araştırmalar Dergisi, 12(4), 1425-1442.

Ekonomi bakanlığı (2016). Serbest Bölgeler, http://www.ekonomi.gov.tr, 21.06.2016.

Gerni, C., Emsen, Ö. S., & Değer, M. K. (2008). *İthalata Dayalı İhracat ve Ekonomik Büyüme: 1980-2006 Türkiye Deneyimi*. Ulusal iktisat kongresi, 20-22.

Granger, C. W., & Newbold, P. (1974). *Spurious regressions in econometrics*. Journal of econometrics, 2(2), 111-120.

Gül, E., & Ekinci, A. (2006). *Türkiye'de Reel Döviz Kuru ile İhracat ve İthalat Arasindaki Nedensellik İlişkisi: 1990-2006*. Dumlupınar Üniversitesi Sosyal Bilimler Dergisi, (16).

Gürbüz, H., & Çekerol, K. (2002). *Reel Döviz Kuru ile Dış Ticaret Haddi ve Bileşenleri Arasındaki Uzun Dönem İlişki*. Afyon Kocatepe Üniversitesi İktisadi ve İdari Bilimler Fakültesi Dergisi, 4(2), 31-46.

Hao, S. (2010). *Study on Relationship Between Trade Development of Waigaoqiao Free Trade Zone and Economic Growth in Shanghai*. In Orıent Academıc Forum, 2(1).

İmamoğlu, İ. K. (2024). *Serbest Bölgelerin Ekonomideki Etkinliği: Türkiye'deki Serbest Bölgeler Üzerine Bir Değerlendirme*. Socıal Scıences Studıes Journal (SSSJournal), 5(50), 6542-6554.

ITO - İstanbul Ticaret Odası (1960), *Transit Ticaretin Memleketimiz Bakımından Ehemmiyeti ve İstanbul'da Bir Transit Antreposu Kurulması*, C. 76, S. 3-4, Mart-Nisan.

Karaçor, Z., & Gerçeker, M. (2012). *Reel Döviz Kuru ve Diş Ticaret İlişkisi: Türkiye Örneği (2003-2010)*. Sosyal Ekonomik Araştırmalar Dergisi, 12(23), 289-312.

Karagöz, M., & Doğan, Ç. (2005). *Döviz Kuru Diş Ticaret İlişkisi: Türkiye Örneği.* Fırat Üniversitesi Sosyal Bilimler Dergisi, 15(2), 219-228.

Kennedy, P., Sarımeşeli, M., & Açıkgöz, Ş. (2006). *Ekonometri kılavuzu*. Gazi Kitabevi.

Li, S., Liu, J., & Kong, Y. (2021). *Pilot free trade zones and Chinese port-listed companies performance: An empirical research based on quasi-natural experiment*. Transport Policy, 111, 125-137.

Öncel, A., & Demirtaş, Ş. C. (2016). Serbest Bölgelerin Dış Ticaret Üzerine Etkisi: Türkiye Örneği (2000-2016). İşletme Öğrencileri Kongresi 3rd, 25.

Öncel, A., & Demirtaş, Ş. C. (2017). *Serbest bölgelerin dış ticarete etkileri: Türkiye üzerine ARDL modeli ile ampirik bir uygulama*. Eskişehir Osmangazi Üniversitesi İktisadi ve İdari Bilimler Dergisi, 12(1), 65-82.

Sun, Y., Jiang, L., Hu, Y., & Yang, J. (2025). *Do pilot free trade zones promote green transformation? Evidence from a quasi-natural experiment*. The article is under review, https://doi.org/10.21203/rs.3.rs-5904894/v1

Tapşin, G., & Karabulut, A. T. (2013). *Reel Döviz Kuru, İthalat ve İhracat Arasındaki Nedensellik İlişkisi: Türkiye Örneği*. Akdeniz İİBF Dergisi, 26, 190-205.

Terzi, H., & Zengin, A. (1999). *Kur Politikasının Dış Ticaret Dengesini Sağlamadaki Etkinliği: Türkiye Uygulaması.* Ekonomik Yaklaşım, 10(33), 48-65.

Toda, H. Y., & Yamamoto, T. (1995). *Statistical İnference in Vector Autoregressions with Possibly Integrated Processes*. Journal of Econometrics, 66(1-2), 225-250.

Tyler, W. G., & Negrete, A. C. A. (2009). *Economic growth and export processing zones: An empirical analysis of policies to cope with Dutch disease*. World Development, 18(2), 220-241.

Yapraklı, S. (2010). *Türkiye'de Esnek Döviz Kuru Rejimi Altında Dış Açıkların Belirleyicileri: Sınır Testi Yaklaşımı.* Ankara Üniversitesi SBF Dergisi, 65(04), 141-163.

Yıldırım, E., & Kesikoğlu, F. (2012). *İthalat-İhracat-Döviz Kuru Bağımlılığı: Bootstrap ile Düzeltilmiş Nedensellik Testi Uygulaması.* Ege Academic Review, 12(2), 137-148.

Yıldırım, S., & Kılıç, E. (2014). *Döviz Kuru Volatilitesinin Türkiye'nin Euro Bölgesi İhracatına Etkisi: Kesikli Dalgacık Dönüşümü ile Panel Veri Analizi*. Atatürk Üniversitesi Sosyal Bilimler Enstitüsü Dergisi, 18(1), 425-440.

Yılmaz, A., & Çapraz, K. (2014). *Comparison of free zones in Turkey by Means of DEA*. Data Envelopment Analysis and Performance Measurement, 77.

Zengin, A., (2001). *Reel Döviz Kuru Hareketleri ve Dış Ticaret Fiyatları (Türkiye Ekonomisi Üzerine Ampirik Bulgular*. Cumhuriyet Üniversitesi İktisadi ve İdari Birimler Dergisi, 2(2), 27-41.